# Living Bento: Heartbeat-Driven Noodles for Enriched Dining Dynamics

Weijen Chen
Keio University Graduate School of
Media Design
Yokohama, Japan
weijen@kmd.keio.ac.jp

Qingyuan Gao
Keio University Graduate School of
Media Design
Yokohama, Japan
girafferyeo13@gmail.com

Zheng Hu
Keio University Graduate School of
Media Design
Yokohama, Japan
jphuzheng@gmail.com

Kouta Minamizawa
Keio University Graduate School of
Media Design
Yokohama, Japan
kouta@kmd.keio.ac.jp

Yun Suen Pai
School of Computer Science
University of Auckland
Auckland, New Zealand
yun.suen.pai@auckland.ac.nz

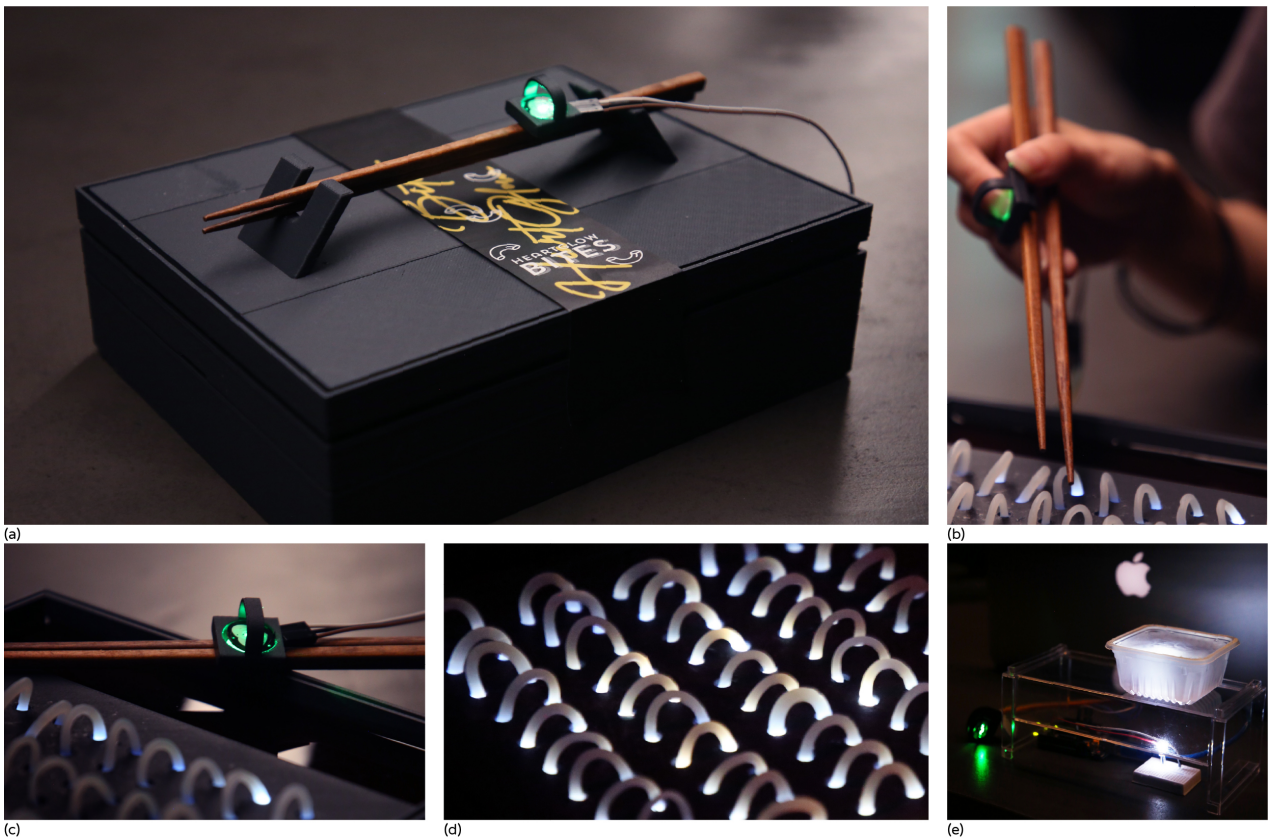

**Figure 1: (a) Living Bento (b) Chopsticks with Pulse Sensor (c) Sauce Compartment (d) Glowing Noodles (e) User Study Setup**

## Abstract

To enhance focused eating and dining socialization, previous Human-Food Interaction research has indicated that external devices can support these dining objectives and immersion. However, methods that focus on the food itself and the diners themselves have remained underdeveloped. In this study, we integrated biofeedback with food, utilizing diners' heart rates as a source of the food's appearance to promote focused eating and dining socialization. By employing LED lights, we dynamically displayed diners' real-time physiological signals through the transparency of the food. Results revealed significant effects on various aspects of dining immersion, such as awareness perceptions, attractiveness, attentiveness to each bite, and emotional bonds with the food. Furthermore, to promote dining socialization, we established a "Sharing Bio-Sync Food" dining system to strengthen emotional connections between diners. Based on these findings, we developed tableware that integrates biofeedback into the culinary experience.

## 1 Introduction

Socrates once said, "drinking is a satisfaction of the want, and a pleasure." [94] Food philosopher and scholar Sarah E. Worth similarly notes, "Eating slowly, savoring, and choosing your favorite foods are all ways not just to stimulate the tongue, but to produce genuine gustatory pleasure." [137] However, modern society, driven by industrialization, has prioritized efficiency through fast food and subconscious, mindless eating, causing us to gradually loose attention to and savoring of the dining experience. Wendell Berry, a renowned pioneer in food ethics and culture and recipient of the National Humanities Medal, supports this view: "industrial eating has become a degraded, poor, paltry thing. Our kitchens and other eating places more and more resemble filling stations." [10] Mindless eating not only means eating without paying attention but also brings unhealthy effects to humans [130]. For example, eating while watching screen-based media can negatively impact our digestive health [4, 96, 97].

In this context, by engaging in immersive dining—consciously and attentively enjoying meals—one can increase attention to eating, enhance the pleasure derived from mealtime, and reduce the unhealthy effects of mindless eating. In the field of Human-Food Interaction (HFI), there has been considerable research on immersive dining. Youssef and Spence [144] utilize an immersive dining concept by incorporating different natural sounds with each course pairing to enhance the gastronomic experiences. Additionally, there are numerous examples of VR applications, such as Crofton et al. [19], who explored the impact of traditional booths and VR contexts, revealing that the VR countryside generated significantly higher hedonic scores for food. And Perez et al. [91], who investigated tasting food paired with an immersive remote setting, creating an innovative concept with potential impact in hospitality and tourism. Furthermore, restaurants employing digital technologies such as visual projections and atmospheric soundscapes, including Ultraviolet by Paul Pairet [9, 52] and Sublimotion by Paco Roncero [78], are classic examples of immersive dining experiences that offer higher entertainment value to consumers.

Moreover, some immersive dining research incorporates the concept of mindful eating [6, 62, 119] to promote intentional eating. Mindful eating encourages individuals to focus on and savor their food, while also recognizing its effects on their bodies [56]. As the concept advocates, Feed the Food Monsters [3] employs an AR game designed to engage diners in proper chewing by incorporating their bodies as part of the gameplay, thereby achieving focused eating. However, Khot and Mueller [56] suggest that VR and AR in mindful eating research could lead to environments where food visibility is compromised. Moreover, most existing solutions are standalone and do not integrate with current eating practices, requiring significant motivation and commitment.

Consequently, our research questions are as follows: **RQ1:** How can we strengthen the connection between individuals and their food in alignment with the core value of awareness in mindful eating, in order to reduce the sense of detachment and standalone experience? **RQ2:** How can we achieve immersive dining and emotional communication at the dining table while clearly presenting the food itself and integrating past dining habits, including communal dining situations? Utilizing biofeedback provided by diners and applied directly to the food may offer a solution.

Deng et al. [22] indicate that embedding human-like traits into food movements can deepen the emotional connection between diners and their dishes, and it can also increase customer loyalty [21]. Biofeedback aligns with the concept of anthropomorphized food interactions, as visualized biofeedback can present human-like traits. Psychologists from Oxford University have also found that arbitrary visual symbols (e.g., circles, squares) can gain special significance through association with oneself due to the self-prioritization effect [109]. In addition to achieving intentional eating by deepening the relationship between people and their food, the unpredictability of the biofeedback provided by diners can also increase the appeal of the food and enhance the immersion in the dining experience. Gaver et al. [37] argue that unforeseen elements in interactive design make systems "mysterious and thus attractive." The study "From Plating to Tasting" [22] also indicates that the unforeseeable dynamics seemed to amplify diners' immersion and potentially deepen their appreciation of the food.

In light of this, we propose a new approach to dining using diners' heartbeat data. The Living Bento is a dining vessel that employs lighting patterns to reflect the diner's real-time heart rate. This design not only enhances the emotional connection between the food and the diner by providing immediate feedback on the diner's physiological state, thereby promoting focused and social eating, but it also makes the food appear to glow due to the bento's structural design. Furthermore, Living Bento integrates sensors into the tableware, allowing users to maintain their existing dining habits without additional learning.

### 1.1 User Study 1 Aims

To address RQ1, we have formulated specific hypotheses regarding dining alone with one's own glowing heartbeat:

(1) H1: Heartbeat-driven glowing food provides greater awareness in the dining experience compared to non-glowing or randomly glowing food.

(2) H2: Heartbeat-driven glowing food offers more emotion-related immersion in the dining experience compared to non-glowing or randomly glowing food.

(3) H3: Heartbeat-driven glowing food delivers a higher level of sensory appeal in the dining experience compared to non-glowing or randomly glowing food.

### 1.2 User Study 2 Aims

To address RQ2, we applied heartbeat-driven glowing food to social dining, referred to as the "Sharing Bio-Sync Food" system, where diners exchange their physiological states with their dining companions, displaying the partner's data in their own food, in order to explore changes in social dining experiences. Our investigation into exchanging glowing heartbeats through food in social dining is driven by the following factors:

(1) Social Connection: Food acts as a medium for social interaction. We envision that sharing each other's heartbeat through food can strengthen social connections and reduce detachment.

(2) Dining Motivation: Numerous studies have shown that social dining can promote food intake [45, 74, 80, 102]. We propose that sharing each other's heartbeat through food may enhance existing dining motivation and appetite.

(3) Innovative Dining Experience: Visually, as the food appears to glow as if it were emitting light on its own, we anticipate that this will increase its appeal to diners, thereby offering a novel dining ritual.

### 1.3 Contributions

Overall, applying biofeedback to food opens a new dimension in the study of HFI, exploring new possibilities to enrich our eating rituals. Our three main contributions are as follows:

(1) We integrate research on the attentiveness aspect of mindful eating and biofeedback within the HCI field, using diners' heartbeat data to strengthen the connection between individuals and their food.

(2) We found that heartbeat-driven food significantly enhanced awareness and appreciation of the food's appearance, the attractiveness and immersion of the dining experience, the emotional bond between the individual and the food, and attentiveness to each bite, with particular significance found in the emotional impact from the food, as supported by two user studies, responding to RQ1.

(3) We not only focus on the impact of heartbeat on individual immersive dining experiences, but also explore its effects on commensal dining, bridging connections between individuals, thus responding to RQ2.

## 2 Related Works

### 2.1 Mindful Eating and Immersive Dining

Mindful eating involves intentionally paying non-judgmental attention to the experience of oneself, food, and the act of eating. It enables individuals to gradually become aware of both internal states (such as thoughts, emotions, hunger, taste, and satiety) and external aspects (such as the nutritional value of various foods) [62]. The core values of mindful eating encompass various aspects, including awareness, distraction, emotional response, external cues, and disinhibition [34]. As Mindfulness-Based Eating Awareness Training (MB-EAT) aims to increase mindful awareness of experiences related to eating and to reduce mindless or habitual reactivity [63], our study will focus on the aspect of awareness. According to the Mindful Eating Questionnaire (MEQ) [34], a quantitative tool used to assess an individual's level of awareness during eating, we will explore various awareness items such as subtle flavors, food appearance, relaxation levels, emotion influence, and food appreciation. Several HFI studies address mindful eating, including Zhang et al. [145], who used a voice assistant to facilitate it, and Parra et al. [89], who explored conversational agents to enhance well-being during mindful eating and cooking. Khot et al. [57] developed SWAN, an augmented spoon to encourage mindful eating, while Epstein et al. [27] created "Food4Thought," an iPhone app for promoting mindfulness.

Immersive dining can enhance dining awareness, aligning with one of the core values of mindful eating. There are various methods to achieve immersive dining, including through different sensory modalities such as visual, auditory, and even tactile experiences. In this study, we focus on the visual aspect, which aligns closely with the theme of luminous food, as lighting not only influences the dining experience but also serves to effectively visualize physiological data. Kantono et al. [51] examined restaurant lighting and its contextual appropriateness, while Bell et al. [8] found that visual features of the dining environment influence dining experiences and food choices. Suk et al. [115] noted that lighting color affects appetite, with yellow stimulating and red and blue discouraging it. Wardono et al. [131] studied how color, lighting, and décor impact sociability, emotion, and behavioral intentions in social dining. Velasco et al. [125] showed that multisensory cues affect whisky perception. In the field of HFI, Jeesan [49] found VR conditions affect immersion and dining experiences. Batat [5] highlighted that AR can influence consumers' restaurant experiences across sensory, affective, behavioral, social, and intellectual dimensions, ultimately enhancing food well-being. Additionally, restaurants like The Fat Duck [110] and The Alchemist [141] use technology for immersion, while Le Petit Chef [5] provides an AR experience in various restaurants.

### 2.2 Commensal Dining

The shared meal is a universal human phenomenon; few things express *companionship* more profoundly—derived from the Latin *cum*, 'together', and *panis*, 'bread'—than the shared meal [109]. Eating alone increases anxiety and is a risk factor for heart disease, diabetes, metabolic syndrome, and loneliness [65]. For men, eating alone and living alone are associated with a higher incidence of abnormal weight [118], and living alone also exacerbates issues related to food waste [132]. Spence [109] notes that evidence from both laboratory and naturalistic dining studies indicates that food consumption typically increases when dining with others. Sobal and Nelson [107] highlight that sharing meals also involves sharing experiences, establishing essential social connections, and cherishing the sense of togetherness. Fischler et al. [33] further confirm that commensal dining within families is a crucial mechanism for fostering togetherness and nurturing familial bonds. R.I.M. Dunbar [25] also points out that eating together triggers the brain's endorphin system, which plays a key role in social bonding [109]. Additionally, the dining experience is closely linked to emotions. Dess et al. [23] reveal that emotion affects taste perception, with individuals experiencing low pleasure rating bitterness higher and sweetness lower after stress compared to their high-pleasure counterparts. Moreover, since olfactory perception significantly influences taste perception[104, 109], Chen et al. [12] find that emotion and personality influence olfactory perception.

Numerous HFI studies focus on enhancing commensal dining. Alhasan et al. [1] explored playful remote digital social eating, while Wang et al. [128] proposed cloud-based dining to balance social interaction with healthy eating. TableTalk [30] turns personal devices into a shared table display to improve interactions. Ferdous et al. [31] used family mealtimes to foster deeper connections through shared content, supporting togetherness and in-depth discussion. Other solutions include Fobo [54], which uses robots as dining companions, and Arm-A-Dine [76], an augmented social eating system with wearable robotic arms. Nabil et al. [79] developed ActuEating, which changes the table's shape and color to boost social engagement.

## 2.3 Personalized and Biofeedback

Due to the self-prioritization effect [114], individuals tend to prioritize personal information during processing, especially when it is self-referenced [99]. Coca-Cola's "Share a Coke" campaign, which replaced standard labels with consumers' names, led to a significant sales increase for the first time in over a decade [109]. Besides the self-prioritization effect, which explains why people favor items related to themselves, the endowment effect [50]—where individuals often demand significantly more to give up an object than they would be willing to pay to acquire it—and the cocktail party phenomenon [16]—where highly relevant stimuli, such as one's own name, can capture attention in a noisy environment—can also explain this preference: people assign greater value to what they own. Additionally, factors influencing people's preferences are also related to the IKEA effect [85]—the increase in valuation of self-made products. Norwegian researchers have found that even with the same meal kit, people rate their own cooking higher than that prepared by others [109].

In addition to the previously mentioned effects of anthropomorphized food [21, 22] and unpredictability [22, 37] that deepen the emotional connection and appeal between diners and food, understanding why people favor items related to themselves helps clarify the effects when physiological signals are applied to food. Since physiological feedback is derived from the diner's own data, it is easier to obtain the diner's awareness. Furthermore, research indicates that digital mindfulness artefacts can be enhanced through biofeedback [124], as biofeedback can concretize hidden bodily functions and provide users with bodily data that is otherwise difficult to notice [39], thereby encouraging self-awareness and self-reflection [133]. Alternatively, for instance, ambienBeat [14] provides closed-loop biofeedback via tactile stimuli based on users' heartbeat rate (HR) to guide the user's HR to resonate with its rhythm, thereby reducing mental stress. Frey et al. [35] also suggest that publicly sharing such private information can enhance interpersonal connections and increase empathy, as exemplified by the FOLLOWMYHEART livestream[1], which broadcast Shia LaBeouf's heartbeat to all followers on a public webpage. This represents an exploration of personal connection and intimacy across distances[2]. In HFI, several studies apply biofeedback. TastyBeats [55] creates a spectacle with sports drinks based on heart rate data, while EdiPulse [53] converts activity data into 3D-printed chocolates. Han et al. [40] used GSR to enhance immersion in mixed-reality dining. Kleinberger et al. [61] use chewing sounds to affect food perception. HAPIfork[3] vibrates to slow down eating speed. Huang et al. [48] integrated an EMG sensor into glasses to measure muscle activity and detect food intake-related events, and Bedri et al. [7] developed a wearable system with various sensors to detect eating moments. Chun et al. [15] created a necklace to detect eating states through head and jaw movements, while Liftware[4] counters hand tremors on the eating process with real-time adjustments.

Research in HFI often relies on additional equipment, with less focus on directly enhancing the dining experience through the food itself. Khot and Mueller [56] also noted that despite the appeal of immersive media, screen-based media remains prevalent during meals. In biofeedback-related eating experience studies, most focus on movements such as chewing or eating gestures to address food intake issues or enhance tasting experience. A smaller number explore the use of heart rate to generate food-related outcomes after the fact, while real-time dynamic physiological states reflected in food remain underexplored. Additionally, while much research explores biofeedback in HCI for mindfulness, there is less focus on using biofeedback to enhance mindfulness eating awareness. Thomaz et al. [121] also highlighted that automatic dietary monitoring technologies perform well in controlled environments but struggle in real-world settings. We propose three hypotheses and key drivers to investigate the impact of heart rate biofeedback lighting patterns on the dining experience, aiming to bridge the gap between equipment, food, and individuals. This approach seeks to improve dining experiences while maintaining feasibility for restaurant settings and to inform the development of dining utensils.

## 3 User Study 1

Before starting the user study, we tested respiratory rate and heart rate as biofeedback sources, selecting them based on their compatibility with tableware for an enhanced dining experience. Although we initially preferred respiratory rate due to its link to mindful eating [62] and awareness of hunger cues [123], preliminary testing showed participants focused more on controlling the food's lighting by controlling their breath, which led to unnatural eating behaviors, contradicting our goal. Therefore, we switched to heart rate data, which, due to its difficulty in human control, led to more focus on eating, better supporting our objective.

---

[1]https://www.dazeddigital.com/artsandculture/article/24085/1/shia-labeouf-and-collaborators-present-followmyheart

[2]https://www.dazeddigital.com/artsandculture/article/24173/1/what-we-learned-from-following-shia-labeouf-s-heart

[3]https://www.cnet.com/reviews/hapifork-bluetooth-enabled-smart-fork-preview/

[4]https://www.liftware.com

In User Study 1, our objective was to test H1, H2, and H3, which suggest that heartbeat-driven illumination in food enhances awareness, increases emotion-related immersion, and improves sensory appeal in the dining experience compared to non-glowing or randomly glowing food.

## 3.1 Taste Stimuli

In this study, we used Kudzu Jelly Noodles (kuzukiri) as our food sample. The primary ingredient is ground arrowroot [41], which is a type of starch gel food [46] and dessert jelly [143]. We chose a translucent food sample to leverage the inherent transparency of the ingredient [105] to enable the food to glow. We selected IAFOODS' Kudzu Jelly Noodles[5] due to its advantages in consistency and lack of color and strong smell. Previous cross-modal correspondence studies have explored the relationship between color and taste [111], and research in neurogastronomy has demonstrated the influence of olfaction on taste [104], so using this sample helps avoid the influence of these additional factors.

## 3.2 Visual Stimuli

The selection of visual stimuli in User Study 1 was inspired by Suk et al.'s [115] work on lighting and food pairing tasks, as well as Armstrong et al.'s [2] comparative testing of biofeedback efficacy, which involved three modes: biofeedback-driven mode, random mode, and no mode as a baseline.

In the biofeedback-driven mode, we used heartbeat detection as the driving signal, utilizing the PulseSensorPlayground.h[6] library with a 500 Hz sampling rate, in analog form, and a range of 0 to 1023 on the Arduino IDE 1.8.18 platform. The LED, controlled via PWM pins, creates a gradient flashing effect: upon detecting a heartbeat, it brightens to maximum and gradually dims. In random mode, the LED flashes at maximum brightness with intervals randomly ranging from 10 to 350 milliseconds. In both the heartbeat-driven and random modes, we used an Arduino UNO R3 as the control board, with the entire system running on a MacBook Pro macOS Mojave. The DiCUNO white LED operates at 3.0-3.3 volts and 20mA, with a maximum luminous intensity of 12000-14000 millicandela and a color temperature of 8000K, indicating a cool light source. The baseline hardware setup was consistent with the above, with the only difference being the absence of illumination.

In summary, we designated the initial baseline mode as Mode A, the random mode as Mode B, and the heartbeat-driven mode as Mode C.

## 3.3 Participants

A total of 59 participants were recruited via social media for the two user studies and a follow-up study. For User Study 1, the power analysis in SPSS indicated a required sample size of 27 participants (Cohen's d = 0.5, power = 0.8, $\alpha$ = 0.05). One participant was excluded due to illness, and since the deviation is minimal, we decided a final sample of 26 participants (19 female, 7 male) aged 16 to 65 years (M = 37.26, SD = 13.43). Prior to participation, each participant provided written informed consent, confirmed no allergies to the ingredients used in the study, and verified normal sensory function in both visual and gustatory aspects. Participants exhibiting symptoms of a cold were excluded. The study received an exemption approval from the university's Ethics Review Board. As a token of appreciation, each participant was given a small dessert (e.g., cake, cookies, or other confectioneries) upon completion of the study.

## 3.4 Procedure

*3.4.1 Procedure Overview.* Participants were required to attend a session lasting 30 minutes, during which they consumed a serving of approximately 55 grams of Kudzu Jelly Noodles for each various lighting patterns. To minimize visual distractions and environmental influences, the experiment was conducted in a uniformly dimly lit space. During the study, participants were guided to complete a series of dining tasks under different lighting patterns. After each task, they were required to scan a QR code linked to an online questionnaire on their personal mobile devices and then self-assess their dining experience. Before testing with randomly ordered lighting patterns, an initial no-light condition was always conducted as the first step in all experiments. According to previous dining experience studies [13, 61], this provided all participants with a baseline practice in the no-light condition, enabling them to become familiar with using the pulse sensor and answering the questions.

The process involves the following steps: 1) starting with the initial mode (A), 2) evaluating their dining experience, 3) presenting the lighting patterns (B/C) in a random order, 4) evaluating their dining experience, and 5) proceeding to the next mode (B/C). This procedure is illustrated in Figure 2:

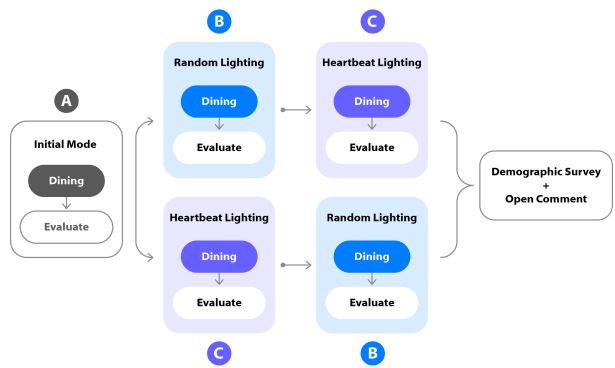

**Figure 2: The Process of User Study 1**

*3.4.2 Assessment.* According to the dining questionnaire design by Winkens et al. [136], adjustments were made to existing validated questionnaires to better align with the research focus. The questions in User Study 1 were inspired by the MEQ [34], which includes dimensions such as awareness of eating, Spence [109]'s advocacy for customized food enhancing preferences, and Deng et al. [22]'s observation that anthropomorphized food fosters emotional connection. The survey evaluated participants' dining experiences through 10 questions, including: awareness of subtle flavors, awareness of the food's appearance, appreciation for the food's appearance, awareness of relaxation during the meal, attentiveness to savoring each bite of food, extent to which the food influences emotions during the meal, overall deliciousness, overall attractiveness, overall immersion, and emotional bond with food. The assessment method was inspired by Chen et al. [13] on dining experience evaluation methods. A 9-point Likert scale was used, with the intensity of noodle tasting experience under different lighting patterns rated from 1 to 9 (1 = not at all, 9 = extremely). The question format was as follows: "Please evaluate your dining experience: awareness of the food's appearance," and participants were asked to provide a numerical response between 1 and 9. Participants took a 5-minute break between modes, and the questionnaire concluded with a demographic survey and an open comment section, allowing participants to provide additional feedback.

To set up the dining experience to simulate the effect of the food glowing by itself, our test setup, as illustrated in Figure 3:

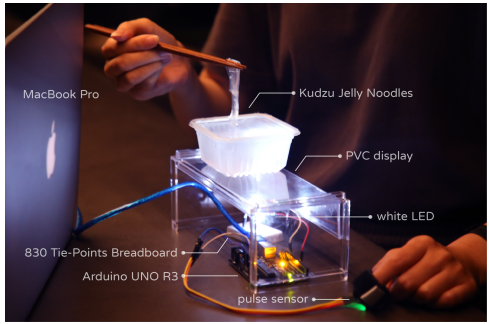

**Figure 3: The Setup of User Study 1**

### 3.5 Results

Based on the following data, our H1, H2, and H3 were assessed in User Study 1, with H2 being fully supported and H1 and H3 being partially supported. Results were analyzed using the Mac version of SPSS 29.0.2.0 (SPSS, Inc.). We conducted a MANOVA and a follow-up t-test analysis to examine the effects of different lighting patterns (no illumination, randomly glowing, and heartbeat-driven illumination) on participants' ratings of their dining experience.

Based on the ratings given to Kudzu Jelly Noodles across ten different assessment items with no illumination, randomly glowing illumination, and heartbeat-driven illumination, a MANOVA revealed a significant overall effect of lighting patterns on the ratings of these dining experiences: food appearance awareness ($F_{(2, 75)} = 14.217$, $p < 0.001$), food appearance appreciation ($F_{(2, 75)} = 12.381$, $p < 0.001$), attentiveness to each bite ($F_{(2, 75)} = 3.967$, $p = 0.023$), emotional influences ($F_{(2, 75)} = 5.604$, $p = 0.005$), overall attractiveness ($F_{(2, 75)} = 15.099$, $p < 0.001$), overall immersion ($F_{(2, 75)} = 7.759$, $p < 0.001$), and emotional bond with food ($F_{(2, 75)} = 4.055$, $p = 0.021$).

#### 3.5.1 The Impact of Heartbeat-Driven Illumination on the Dining Experience. Figure 4 shows the evaluation ratings for the noodles under the three lighting modes. Compared to the baseline (no illumination), we observed significant effects of heartbeat-driven illumination on the ratings for food appearance awareness (95% CI [1.71, 3.29], $t(25) = 6.56$, $p < 0.001$, Cohen's $d = 1.94$), food appearance appreciation (95% CI [1.66, 3.57], $t(25) = 5.63$, $p < 0.001$, Cohen's $d = 2.37$), attentiveness to each bite (95% CI [0.41, 2.67], $t(25) = 2.80$, $p = 0.010$, Cohen's $d = 2.80$), emotional influences (95% CI [0.89, 3.11], $t(25) = 3.72$, $p = 0.001$, Cohen's $d = 2.74$), overall attractiveness (95% CI [1.67, 3.25], $t(25) = 6.39$, $p < 0.001$, Cohen's $d = 1.96$), overall immersion (95% CI [0.87, 3.13], $t(25) = 3.64$, $p = 0.001$, Cohen's $d = 2.80$), and emotional bond with food (95% CI [0.63, 2.83], $t(25) = 3.24$, $p = 0.003$, Cohen's $d = 2.72$). Overall, heartbeat-driven illumination significantly enhanced the intensity of food appearance-related aspects (awareness and appreciation), overall experience (attractiveness and immersion), emotional aspects (emotional influences and emotional bond), and attentiveness to each bite.

#### 3.5.2 The Impact of Randomly Glowing Illumination on the Dining Experience. Figure 4 shows the evaluation ratings for the noodles under the three lighting modes. Compared to the baseline (no illumination), we observed significant effects of randomly glowing illumination on ratings for food appearance awareness (95% CI [1.25, 2.90], $t(25) = 5.20$, $p < 0.001$, Cohen's $d = 2.04$), food appearance appreciation (95% CI [0.82, 2.95], $t(25) = 3.66$, $p = 0.001$, Cohen's $d = 2.63$), overall attractiveness (95% CI [0.89, 2.88], $t(25) = 3.89$, $p < 0.001$, Cohen's $d = 2.47$), and overall immersion (95% CI [0.45, 3.01], $t(25) = 2.79$, $p = 0.010$, Cohen's $d = 3.17$). Overall, randomly glowing illumination significantly enhanced the intensity of food appearance-related aspects (awareness and appreciation) and overall experience (attractiveness and immersion).

#### 3.5.3 Qualitative Analysis. In User Study 1, we conducted a qualitative analysis of 49 spontaneously open comments to elucidate the impact of visual stimuli on the dining experience. These comments were responses to the final open-ended question, "Any other comments?" The transcribed textual records were coded and categorized by both the researcher and the second author, resulting in four themes.

**Attention (12 cases, =24.49%).** In addition to the general comments from P3, P6, P12, and P19, which stated that "*eating glowing food is cool, fascinating, and interesting,*" P5, P16, P18, P21, and P26 noted that "*when the lighting response was applied to the food, it enhanced the focus, making it easier to notice and observe the food's appearance, thus increasing its attractiveness.*" P14 specifically mentioned, "*I noticed that my chewing frequency increasingly synchronized with the light flashes, which was a truly enjoyable experience!*" However, P13 provided a differing comment, stating, "*The flickering lights still distracted from the taste experience.*" P25 provided distinct feedback for the heartbeat-driven and random modes: "*Mode B [random] made tasting more focused, while Mode C [heartbeat] provided the most comfortable overall flashing intensity.*" Most comments regarding attentiveness aligned with the findings from the quantitative analysis.

**Emotional Impact (17 cases, =34.69%).** Many participants provided varying evaluations of the B (random) and C (heartbeat)

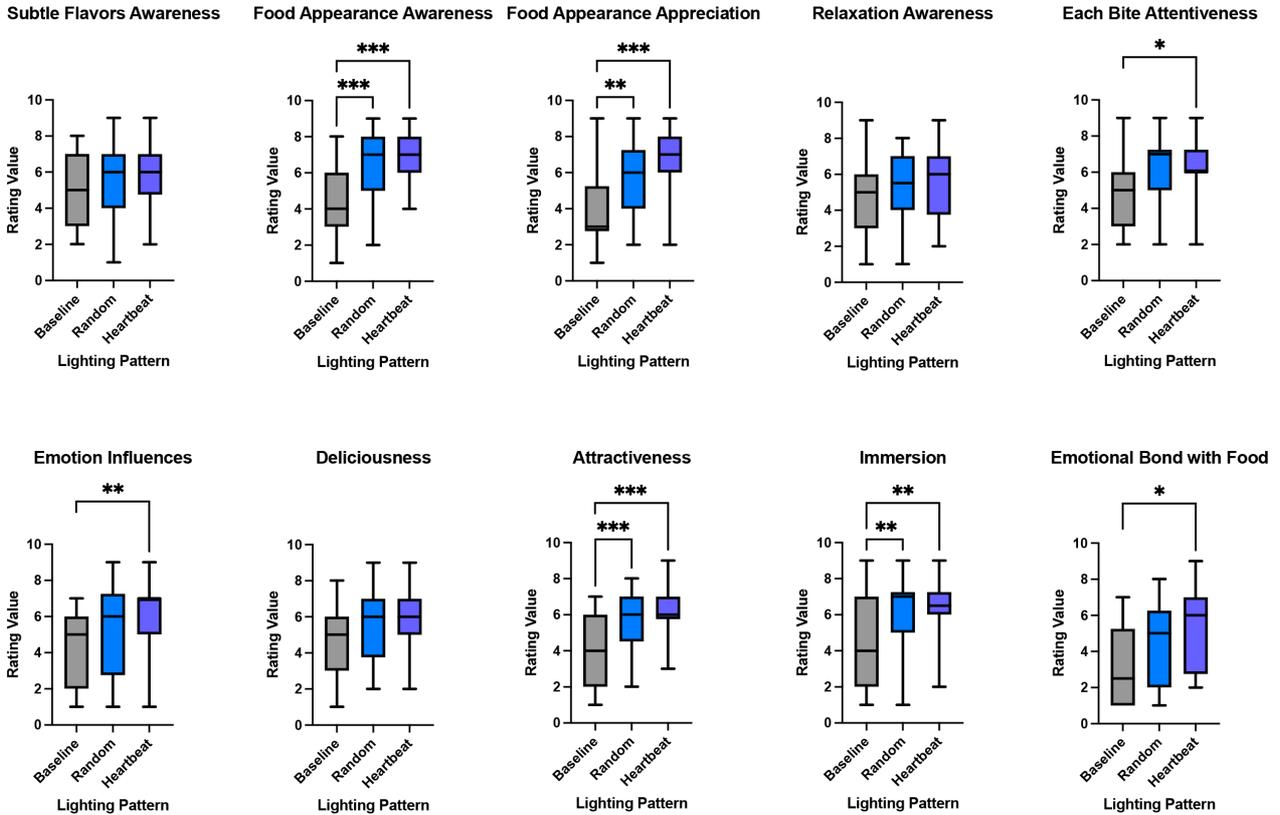

**Figure 4: Dining Experiences Ratings Influenced by Lighting Patterns**

modes. For the random mode, positive feedback included comments such as P1, P16 and P19's observation that "*it created a sense of excitement and unconsciously increased interest in the food,*" P15's view that "*the atmosphere has a sense of mystery; it evokes a festive feeling akin to Christmas or Valentine's Day.*" Conversely, several participants expressed negative feedback, such as P4's feeling that it was "*a bit irritating,*" P7 and P8's observation that it made them "*unable to relax and felt tense with some appeal of pressure during eating,*" P9, P10 and P13's concern that "*it distracted from the taste experience by focusing too much on the visual changes.*"

Regarding the heartbeat mode, positive evaluations included P1's comment that "*the steady pace aroused curiosity,*" P4's remark that it elicited "*a more appreciative emotion,*" P9 and P19's observation that "*it had less emotional fluctuation but provided a greater sense of immersion,*" and P10, P11, and P25's consensus that "*Mode C offered the most comfortable experience, being a pleasant and relaxing process.*" Additionally, P16 speculated, "*I surmise that Mode C is related to the heartbeat, which enhances the appeal of the immersive experience,*" and P20 noted that "*the lighting significantly affected appetite, particularly when the light flashed in sync with the heartbeat, which was fascinating!*" However, some participants provided less favorable feedback, such as P15's observation that "*Mode C felt like dining under a stable light source, which was better than no*

*light, but lacked stimulation,*" and P18 and P22's concern that "*felt somewhat tense when glowing in sync with the heartbeat.*" Overall, comments from participants indicated that heartbeat-glowing food had a notable impact on emotional responses, with the illumination state affecting the emotional bond between individuals and the food to some extent.

**Eating Speed (6 cases, =12.24%).** Some participants indicated that the frequency of the illumination affected their eating speed. For example, P1, P7, and P14 noted that "*a faster flicker rate created a sense of excitement and made it harder to relax, accelerated eating speed,*" and P10 even mentioned, "*Mode B [random] made me want to finish eating quickly.*" P26 commented on both the baseline and heartbeat modes and shared different insights compared to other participants, stating, "*Mode A [baseline] did not encourage slow chewing without special lighting effects, whereas Mode C [heartbeat] allowed for easier observation of the food's appearance, with the light on the small bubbles being more noticeable, resulting in slower eating.*" Comments on eating speed were an unexpected finding, so we plan to include discussions about eating speed in the interview section of the next workshop.

**Flavor Variation (12 cases, =24.49%).** Participants also evaluated the taste and texture of the food under the baseline, random,

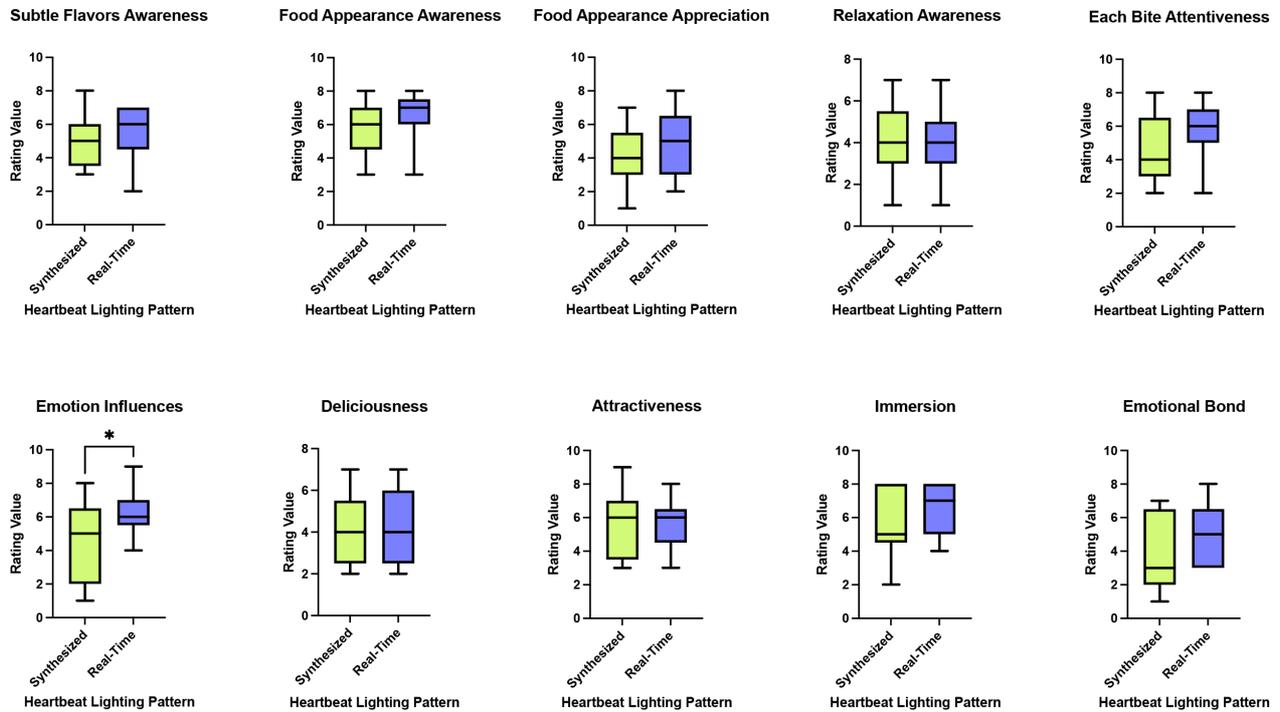

**Figure 5: Dining Experiences Ratings Influenced by Synthesized and Real-Time Heartbeat-Driven Lighting Patterns**

and heartbeat modes. Regarding the baseline mode, feedback included P9's comment: "*It was easier to perceive changes in the food's flavor, but without illumination, the overall experience felt somewhat lacking,*" and P14, P15, P16, and P25 all noted: "*Under Mode A, there was little appeal and emotional variation, but since the food itself had no color, the initial sweetness was particularly surprising, though later on it became less noticeable and simply tasted sweet, feeling like just eating food.*"

For the random mode, feedback included P13's comment: "*Compared to Mode A [baseline], it is harder to experience the food's flavor.*" Positive feedback included P15's: "*The sweetness seems best in Mode B,*" and P25's: "*It allows for tasting fruity aromas.*" Regarding the heartbeat mode, P25 stated: "*It has the most pronounced fruity aroma among the three modes,*" and P26 noted: "*The continuous illumination makes it easier to notice the food's appearance while also experiencing its flavor and texture.*" In contrast, P13 had a different view: "*Mode C is more subdued than Mode B [random], thus experiencing more flavor and concentrating better on eating, but compared to Mode A [baseline], the flickering light still disrupts the taste perception.*" In response to P13's viewpoint, we plan to include a discussion on eating disruption issues in the interview section of the next workshop.

Lastly, some participants provided overall evaluations of the illuminated food, such as P1's comment: "*The fast flicker rate made me feel excited and even enhanced the food's texture,*" while P2, P4, P16, and P22 noted: "*It was a unique experience with minimal impact on taste, but a significant effect on emotions!*"

For User Study 1, both P3 and P9 provided suggestions regarding visual stimuli, stating "*since the light source was only white and the food itself was quite transparent, it is suggested to consider different colored light sources or different food combinations.*" However, since light color can affect the dining experience [115, 131], to minimize variables, we will continue using the same white LED lighting for visual stimulation in the next workshop.

## 3.6 Follow-Up Study on the Effectiveness of Heartbeat-Driven Glowing Food

After finding that heartbeat-driven glowing food enhances the dining experience, we conducted a follow-up user study with 13 participants (9 female, 4 male, aged 22–27, M = 24.92) for further clarification. This study followed the same procedure, with the difference being that it compared synthesized (using pre-recorded heartbeat data from the second author, ranging from Heartbeat #1: 0.000 seconds to Heartbeat #180: 124.079 seconds) and real-time heartbeat-driven lighting patterns. MANOVA revealed a significant effect of heartbeat-driven illumination on emotional influences ratings ($F(1, 24) = 4.554, p = 0.043$). As shown in Figure 5, while no significant effects were observed for food appearance awareness (real-time: M = 6.54, SD = 1.39 vs synthesized: M = 5.92, SD = 1.50) and appreciation (real-time: M = 4.77, SD = 1.92 vs synthesized: M = 4.00, SD = 1.73), attentiveness to each bite (real-time: M = 5.54, SD = 1.85 vs synthesized: M = 4.85, SD = 1.91), immersion (real-time: M = 6.46, SD = 1.39 vs synthesized: M = 5.77, SD = 1.92), and emotional bond with food (real-time: M = 5.00, SD = 1.78 vs synthesized: M =

4.08, SD = 2.33), the real-time pattern consistently yielded higher ratings.

Overall, User Study 1 confirmed the positive impact of heartbeat-glowing food on dining experiences, including food appearance awareness and appreciation, attentiveness, emotional influences, attractiveness, immersion, and emotional bonding with food, compared to the baseline. Notably, emotional influences were more pronounced compared to the synthesized pattern. Building on these findings, User Study 2 further explored the effect of transitioning from solo to social dining.

## 4 User Study 2

The goal of this study is to explore the potential of heartbeat-driven illumination in commensal dining—particularly when diners exchange their glowing heartbeats with their dining companions, a concept referred to as "Sharing Bio-Sync Food." We aim to uncover what new insights this interaction can bring to commensal dining. To achieve this, User Study 2 will be conducted as a workshop to gain a deeper understanding of participants' dining experiences.

### 4.1 Taste Stimuli

In the workshop, the taste stimuli will remain consistent with those used in User Study 1, specifically IAFOODS' Kudzu Jelly Noodles. However, unlike in User Study 1, where participants were prohibited from using the accompanying brown sugar sauce to reduce variables in the quantitative study, this workshop will encourage participants to freely add sauces to better simulate real dining conditions. Additionally, based on feedback from User Study 1, where P3 and P9 noted that the visibility of the translucent ingredients was somewhat insufficient, we will incorporate this suggestion by adding sauces to slightly enhance visibility.

### 4.2 Visual Stimuli

The visual stimuli in this workshop are consistent with the heartbeat-driven mode used in User Study 1, utilizing the same hardware and software. The only difference is that in User Study 2, the food presented to participants will display the glowing heartbeat of their dining partners, rather than their own.

### 4.3 Participants

We recruited 20 participants (14 female, 6 male) via social media, with ages ranging from 19 to 34 years (M = 26.35, SD = 2.90). For this workshop, participants were required to bring a companion, resulting in each session being conducted with pairs, totalling 10 pairs. Prior to participation, each participant provided written informed consent, confirmed no allergies to the ingredients used in the study, and verified normal sensory function in both visual and gustatory aspects. Participants exhibiting symptoms of a cold were excluded. The study received an exemption approval from the university's Ethics Review Board. As a token of appreciation, each participant was given a small dessert (e.g., cake, cookies, or other confectioneries) upon completion of the study.

### 4.4 Procedure

*4.4.1 Procedure Overview.* Each pair of participants was required to attend a 45 minute session (approximately 15 minutes for dining and 30 minutes for post-meal interviews). During this period, each participant consumed approximately 115 grams of an unopened portion of Kudzu Jelly Noodles. To minimize visual distractions and environmental influences, the experiment was conducted in a uniformly dimly lit space.

During the study, participants were guided to sit facing each other for the dining task. Once settled, they were instructed to use a pulse sensor while interacting with the noodles. In the initial approximately three minutes, participants were asked to start eating without the food glowing, allowing them to acclimate to the dining setup. Following this, heartbeat detection was activated, and the phase of "Sharing Bio-Sync Food" commenced. Participants were encouraged to converse freely, maintaining their usual dining habits. They were also encouraged to interact with the noodles according to their preferences and to add the brown sugar sauce provided in the noodle packaging as desired.

*4.4.2 Assessment.* In the post-meal phase, participants from the same pair were invited to participate in an interview to discuss their experience with "Sharing Bio-Sync Food." We used semi-structured interviews to allow participants to elaborate on any relevant or unrelated thoughts, aiming for a deep understanding of the communal dining experience of sharing heartbeats. We prepared ten questions, inspired by feedback from User Study 1 participants, the Social Connectedness Scale (SCS) [66], the Mindful Eating Questionnaire (MEQ) [34], and "From Plating to Tasting [22]":

(1) How does the food glowing with the other person's heartbeat affect your eating speed?
(2) How does the food glowing with the other person's heartbeat affect your experience of each bite?
(3) How does the food glowing with the other person's heartbeat affect your emotions?
(4) How does eating food that glows with the other person's heartbeat affect your feelings towards them?
(5) How do you feel about your involvement with the other person?
(6) How has your dining experience changed over time?
(7) How is your motivation for dining?
(8) How do you feel about the relationship between yourself, the other person, and the food?
(9) How do you perceive this food intake experience compared to the traditional way of eating?
(10) Any other comments?

### 4.5 Results

We collected data consisting of audio and video recordings from all complete workshop sessions, including the period from the beginning of the meal to the end of the interview. We captured all participant conversations, dining behaviors, and group discussions. All 268 comments made during the interviews were transcribed and stored in TurboScribe[7] for management. The interviewer and the second author categorized the transcripts into similar codes, grouped the similar codes together, and analyzed them to form cross-cutting themes. Ultimately, five themes were distilled under the review of the rest of the co-authors.

---

[7]https://turboscribe.ai

*4.5.1 Theme 1: Sensory Appeal of Food (38 cases, =14.18%).* **Visual Attractiveness.** Most participants appreciated the dynamic appearance of the food, citing the interaction between the ingredients and lighting. For instance, P1, P9, P15: "*Even though the food is ordinarily commonplace, adding the glowing dynamic effect makes it more attractive than static food.*" Additionally, P3, P4, P10, P13, P14, and P15 mentioned: "*The food's transparent material allows for optical interaction [...]. It is easy to focus on the food's appearance!*" However, several participants indicated minimal impact, such as: P2, P5, P6: "*I focus more on friends. The steady flashing frequency makes it easy to become background lighting.*" P3 also noted: "*Glowing introduces an additional distraction.*"

**Flavor.** Participants had mixed views on the glowing's impact on flavor. P3 and P4 stated, "*The food tastes sweeter without the light,*" while P9, and P20 remarked, "*I experienced an acidity once the light was on.*" P4 even added, "*It seemed to taste better as I ate more.*" Conversely, P1, P2, P10, P11, P17, P18, and P19 consistently indicated, "*There is not much difference in taste between the glowing and non-glowing conditions.*"

**Food Interaction and Awareness.** Some participants chose to interact with the food, which enhanced their focus while eating. For example, P2: "*I observe the shape of the food, find it enjoyable and interesting, and feel inclined to play with it!*" P4, P6, P13, and P14: "*I used to eat mindlessly, but now I pay attention to the food even while chatting.*"

Overall, most participants viewed the heartbeat-glowing food positively, finding it visually appealing and attention-grabbing. Some focused more on their companions, seeing the heartbeat-driven glow as ambient lighting, while one found it distracting. Opinions on flavor varied, but all praised the playful interaction for enhancing focus. These results align with User Study 1, showing that glowing heartbeat food improved appearance awareness, appreciation, and overall attractiveness compared to the baseline.

*4.5.2 Theme 2: Human Interaction (85 cases, =31.72%).* **Emotion Recognition.** Most participants viewed the food as an additional channel for understanding others' emotions. P1, P9, P10, P15, P19, and P20: "*Since heartbeats cannot be disguised, it provides an extra reference point. Even if the other person is expressionless, you can still sense their true emotions.*" Additionally, P1 and P20 noted: "*The food's glow, with no fixed pattern, leads to variations in the flashing effect depending on the conversation. For instance, when discussing a very angry topic, the light flashes more rapidly.*" Some dialogues also highlighted this:

> P9: "*I noticed you like cute European guys and otaku culture! Because your heartbeat started to accelerate!*"
> P10: "*You figured it out [laughter].*"
> P9: "*Haha, captured some hidden messages [laughter].*"

However, P13, P17, and P18 felt the glowing food did not accurately convey emotions, noting, "*When the flashing is very rapid, it's unclear whether the other person is nervous or simply deeply engaged in the topic.*"

**Behavior and Emotion Understanding.** Some participants were interested in comparing heartbeats with others. For example, P1, P5, P11, P19 stated, "*I became more engaged and tried to sense the differences between my heartbeat and theirs. Sometimes, I observed our heartbeats synchronizing.*"

Others attempted to control or influence heart rates. P3 and P5 tried to synchronize their heartbeats with their partner's, saying, "*While eating, I tried to align my heartbeat with theirs and establish a connection, but I found I could only maintain it for a short period.*" Conversely, P16 and P18 wanted to influence their partner's heartbeat: "*I wanted to say things to either speed up or slow down their heartbeat [laughter].*"

Some reflected on their own behavior or sought to understand others. P6 and P16 said, "*When the other person's heartbeat is fast, I become concerned about their well-being. I then reflect on what I might have said to alter their heartbeat, hoping I didn't say something that upset them.*" Meanwhile, P6, P11, and P18 stated, "*If the other person's heartbeat is fast, I find it interesting and want to know what they are thinking.*"

Finally, many participants found heartbeat-glowing food to be an effective communication medium. P9, P10, P13, P14, P15, and P16 noted, "*Compared to just seeing the other person's glowing heartbeat, the food as a medium makes the communication more natural.*"

Overall, most participants viewed heartbeat-glowing food positively as an additional channel for conveying emotions and messages during meals. While three participants found the signals inaccurate, many subconsciously observed heart rate differences, attempted to control them, or reflected on their own or others' states. Notably, heartbeat-glowing food was widely seen as an effective communication medium in communal dining.

*4.5.3 Theme 3: Sensory and Emotional Modulation (95 cases, =35.45%).* **Emotion Influence.** Some participants felt the heartbeat-glowing food enhanced emotional connections. P2, P5, P9, and P15 stated, "*Consuming the heartbeat of another person seems to suddenly bring our relationship to a more intimate level.*" "*It creates a deeper emotional connection, allowing for greater immersion in the conversation and bringing people closer together.*" "*It enhances empathy, allowing me to immediately sense the presence of the other person.*" P14 even mentioned a feeling of companionship: "*It means someone's heartbeat is here, making me feel less lonely.*"

Regarding involvement, P16 and P20 said, "*It feels as if the other person is participating in my experience [...] since my food state changes based on the other person, it makes me more involved with the entire dining table and environment.*"

Regarding self-emotion, P3, P15, and P20 stated, "*When the other person's heartbeat is slow, I feel more relaxed.*" P15 and P16 said heartbeat-glowing food alleviated awkwardness: "*Without a topic or medium, there is often an awkward silence between two people. However, having glowing food as a medium makes it more relaxed and diverts attention.*" Some also found it romantic. P11 and P15 noted, "*It feels like a romantic atmosphere, like during a date, and eating the heartbeat has a magical feeling.*" The dialogues also reflected a romantic ambiance:

> P9: "*Wow, how ambiguous! I think I'm falling for you [laughter].*"
> P10: "*Emotionally, I'm not falling for you [laughter], but the visual experience makes me happier than usual dining!*"

Regarding the overall experience, P2, P3, P5, P6, P11, P12, P13, P14, and P19 said, "*Every bite is illuminated, it adds an element of playfulness!*" "*It doesn't disturb the meal at all. It increases enjoyment* *and is good for conversation.*" Others reported feeling excited, anxious, or awkward. P3, P4, and P11 said, "*It made me excited, with a sense of urgency.*" P19 noted, "*Since it feels like revealing something personal, it adds a bit of nervousness.*" P1 and P4 noted, "*Eating the heartbeat of another person feels somewhat awkward.*"

Some participants noted that familiarity or stable emotions reduced the impact. P3, P4, P5, P6, P7, P8, P12, P13, and P14 mentioned, "*We are too familiar with each other, so it might work better on a date, but we are friends.*" "*Since our emotions are stable, it became a monotonous tone. If we had an argument, it might be more interesting.*"

Finally, heartbeat-glowing food increased dining motivation for most participants. P1, P2, P9, P10, P11, P15, P19, and P20 noted, "*When the other person is particularly funny, I want to eat a few more bites.*" "*It feels novel and visually appealing, providing motivation.*" P7 and P18 mentioned that even if they didn't like the food, the system encouraged them to eat more: "*Without the light, I might not eat at all, but with this device, I am more inclined to eat unpleasant food. For example, if I were given medicine, I'd be more willing to take it with this mechanism.*" Only P12 stated, "*There is no additional motivation.*"

**Physiological Influence.** Most participants discussed the effect on eating speed. Some, like P4 and P5, said, "*I ate a bit faster than usual.*" Others, such as P9, P10, and P20, shared the opposite experience: "*The eating speed slowed down because the atmosphere was quite pleasant, so I wanted to savor each bite a bit longer.*" P17 remarked, "*I ate in sync with the rhythm of the light.*" However, most participants, including P1, P3, P6, P8, P15, P16, and P19, noted, "*I felt that the steady flickering had minimal impact on the eating speed.*" Some mentioned other physiological effects. P4 stated, "*I felt that the left side of my body was more stimulated, and my heartbeat seemed to quicken.*" P3 added, "*I felt my heartbeat accelerate.*"

**Illusion.** Interestingly, P4, P14, and P19 reported experiences related to illusion. P4 stated, "*I felt that the glowing food matched my own heartbeat, creating a sense of connection. Sometimes, if it beat very quickly, I would think it was my own heartbeat, making me feel more excited.*" P14 mentioned, "*It's like having someone with me, even though the person is not physically present.*" P19 expressed, "*It felt like I was consuming their [dining companion's] food.*"

Overall, most participants felt that "Sharing Bio-Sync Food" enhanced emotional connection, influenced their emotions, and made dining more enjoyable, with some describing it as romantic or ambiguous. Some experienced excitement, nervousness, or awkwardness, while others found it alleviated awkwardness. Familiarity with dining partners reduced emotional impact. Most participants said the system motivated them to eat disliked foods, though one noted no additional motivation. A few adjusted their eating speed to match their partner's heartbeat, but many observed minimal impact. Some mistook their partner's heartbeat for their own or perceived the food as a substitute for their partner. Two participants reported increased heartbeats and feeling more excited in response to their partner's heartbeat.

*4.5.4 Theme 4: Awareness Duration (17 cases, =6.34%).* Participants reported that the heartbeat-glowing food felt novel at first but gradually blended into the background. P1, P2, P11, P13, P14, and P19, noted, "*Initially, it felt more novel and interesting, so I paid attention to it. However, as the eating process progressed and the light frequency remained stable, I became accustomed to it, although occasionally I would still notice it.*" Some participants noted the flickering light did not affect their eating. P14 and P15 stated, "*It's very stable, so I get used to it; I don't think it's annoying.*"

Overall, most participants gradually adapted to the glowing's stable frequency, perceiving it as ambient lighting, with some noting no disruption to eating. Future research could examine the persistence of these effects.

*4.5.5 Theme 5: Suggestions (33 cases, =12.31%).* At the end of the interviews, participants provided valuable feedback in response to the concluding question, "Any other comments?"

**Data Readability.** Many participants noted that interpreting emotions from the flickering light required additional analysis. For example, P2, P5, P6, P9, P11, P12, P14, and P17 reflected this issue: "*The data presentation lacks detail, as it requires personal interpretation and analysis. Data visualization would be beneficial. The signals should be processed differently to make them more comprehensible.*"

**Device Productization.** Participants suggested improving device usability by encasing components like the Arduino to reduce distractions and enhancing comfort with the pulse sensor. P5 and P15 said, "*The equipment can be distracting; it would be better to encase the devices.*" P11, P17, and P19 added: "*The fixed setup feels a bit restrictive.*"

**Ingredient Recommendations.** Some participants recommended ingredient choices. P9 mentioned, "*Switching to other transparent foods is feasible; the glowing food system works well!*" P5 suggested, "*I think this system is suitable for desserts, as desserts are often part of afternoon tea, where the main purpose is social interaction. In contrast, savory dishes or main courses primarily serve to satisfy hunger.*"

Finally, P13 noted, "*If there is another outbreak of a pandemic and people need to be isolated, this system would be ideal for remote dining!*" P14 also highlighted its value in offering companionship during solo dining.

## 5 Discussion

### 5.1 User Study 1 Interpretation

In User Study 1, we tested mindfulness eating awareness, focusing on subtle flavors, food appearance awareness, food appearance appreciation, relaxation, and attentiveness to each bite. H1 was partially supported, with three out of five aspects showing significant differences under heartbeat-driven glowing food conditions: food appearance awareness, food appearance appreciation, and attentiveness to each bite. Although food appearance awareness and appreciation were also significant in the random mode, Figure 4 shows that ratings for the heartbeat mode were more concentrated at higher scores. This suggests that while glowing effects enhance visual appeal, biofeedback signals may resonate more, as reflected in the higher heartbeat mode scores. However, for subtle flavors awareness and relaxation awareness, no significant improvement was observed in either random mode or heartbeat mode. In the subtle flavors awareness test, despite some participants (e.g., P15 and P25) reporting changes in flavor perception, most indicated that lighting patterns had little effect on taste. This result confirms the minimal impact of lighting patterns on taste perception and supports crossmodal research linking color and taste. Previous studies show that white is associated with salty flavors [86, 111, 127]. Since we used white LED lights and sweet-and-sour Kudzu Jelly Noodles (which contain high fructose corn syrup, konjac powder, kudzu powder, gelling agents, and acidifiers), white light did not enhance the perception of sweetness or sourness. Regarding relaxation awareness, participants exhibited varied responses. P10, P11, and P25 found that heartbeat mode provided a comfortable, pleasant, and relaxing experience. Previous studies show that relaxation is often associated with lower blood pressure or heart rate [20]. In contrast, P18 and P22 reported tension, typically associated with increased blood pressure or heart rate [146]. Given the potential for physiological feedback to evoke emotional resonance [14] and participants' self-reported feelings, we infer a connection, supported by research suggesting that biofeedback can regulate physiological states and emotions [18].

In the emotion-related immersion tests, we assessed the extent to which food influenced participants' emotions, emotional bond with the food, and overall immersion. H2 was fully supported, with only heartbeat-driven glowing food showing significant effects. This supports previous HCI research suggesting that anthropomorphized food can evoke emotions [59, 95, 122] and confirms the impact of anthropomorphized food on the dining experience.

Regarding sensory appeal, we tested overall attractiveness and deliciousness based on the self-prioritization effect theory [114]. H3 was partially supported. Both random and heartbeat modes significantly affected overall attractiveness (Figure 4), with heartbeat mode showing higher scores. This suggests that while glowing effects enhance attractiveness, biofeedback signals might have a more noticeable influence. However, in the overall deliciousness test, although heartbeat mode ratings were higher than baseline and random modes, no significant differences were found. This may be due to the bland taste of the food and the lack of strong visual contrast with the white, semi-transparent glowing effect. While some participants reported the food seemed tastier under different lighting, most indicated little difference across the three modes.

Lastly, a follow-up test confirmed that the real-time heartbeat pattern's effectiveness is not solely due to its life-like nature, showing significant effects on emotional influences compared to the synthesized pattern. This strengthens the earlier finding that H2 was fully supported. Although other items did not show significant effects, the real-time heartbeat pattern received higher ratings for food appearance awareness and appreciation, attentiveness to each bite, immersion, and emotional bonding with food.

## 5.2 User Study 2 Interpretation

As observed in User Study 2, participant feedback generally reflected the results from User Study 1. Aside from similar feedback on the visual appeal and attentiveness towards glowing food, this study focuses on the concept of "Sharing Bio-Sync Food" and explores the relationships between diners, their partners, and the food, as well as the impact of shared heartbeat signal.

Feijt et al. [29] found that intimacy, connectivity, and shared experiences can be enhanced by social biofeedback systems. In User Study 2, many participants indicated that heartbeat-glowing food fostered connection, offering an additional communication channel and strengthening emotional bonds. This supports Khot and Mueller's [56] notion that eating is a form of "social ritual" and Moge et al.'s [77] findings that sharing physiological data can promote emotional convergence and physiological synchrony, key elements for enhancing empathy and social connection. Several participants reported that heartbeat-glowing food allowed them to discern hidden messages from their partners, indirectly validating previous research that unfiltered emotional sharing enhances emotional intimacy through mutual trust [47, 71, 103, 135]. However, some participants in User Study 2 felt unable to accurately interpret their partner's emotional state based on the corresponding glowing frequency. Additionally, participants experienced various emotions such as pleasure, fun, romance, ambiguity, excitement, nervousness, and awkwardness while sharing each other's heartbeat. Some participants found that heartbeat-glowing food reduced awkwardness and facilitated conversation. Sharing biofeedback signals can stimulate new dialogues [24, 42, 67, 69, 70, 90, 126] and foster spontaneous interactions with strangers [28, 73, 77, 98]. However, the emotions elicited were influenced by the level of familiarity between participants. In close relationships (e.g., friends), heightened emotions like romance and excitement were less pronounced. Previous research shows that different interpersonal relationships require various communication channels to achieve varying levels of electronic intimacy [68, 77]. Additionally, the concept of sharing each other's heartbeat seemed to foster empathy, with participants viewing the food as a proxy for their partner, alleviating loneliness in solo dining. In both User Study 1 and 2, only one participant in each study noted the lighting patterns as a distraction. Most participants saw the heartbeat-glowing food as ambient lighting and reported emotional stabilization, shifting from novelty and excitement to acclimatization.

We found that "Sharing Bio-Sync Food" not only influenced emotions but also affected dining behavior. While some participants disliked the Kudzu Jelly Noodles, those who did still ate more due to the system's influence, supporting research on increased food intake in social dining settings [45, 80]. Furthermore, physiological data, including socially relevant information, can aid individuals in making psychological inferences [120]. Many participants in User Study 2 reported that the system prompted them to infer their partner's emotional state and consider how their comments affected it. This aligns with prior research indicating that participants often reflect on emotional causes in social biofeedback systems [106, 138] and may adjust behavior to avoid tension [100, 101]. Additionally, some studies found that users may feel concerned about their partner's physiological function [38], as seen when participants in User Study 2 concerned about the rapid frequency of the glowing food, attempting to control the partner's emotions through conversation. Others, without self-reflection or attempts to control, subconsciously compared their heart rate to the food's glowing frequency. Previous research suggests that shared physiological data stimulates comparative judgments [77, 108] and serves as self-emotion validation [92].

Lastly, the phenomenon of synchronizing [112, 113, 116] or imitating [35, 58, 64] physiological responses led some participants to perceive their heart rate and the glowing food as being in sync, with some attempting to control their own heart rate to match the glowing food's frequency. Overall, in User Study 2, while a few participants noted that their eating speed was influenced by their partner's heart rate, most maintained their usual pace, likely due to their partner's stable emotions.

## 5.3 Prototype Design: Utensils for Heartbeat-Driven Glowing Food

Based on feedback from P2, P5, P6, P9, P11, P12, P14, P15, P17, and P19 in User Study 2, we improved the equipment used in user testing and designed a dedicated prototype, "Living Bento." Living Bento exemplifies the system, integrating traditional aesthetics with modern mechanical engineering and drawing inspiration from the traditional Japanese lunch box, "Jubako[8]".

*5.3.1 Mechanical Design.* The Jubako's multi-layered design suits our needs for separating food from the device and concealing components such as the Arduino UNO R3, breadboard, LEDs, and power supply. It includes three layers (Figure 6): a. Device Layer, b. Isolation Layer, and c. Food Layer, topped with a lid to form a complete lunch box. This structure combines aesthetic concealment with structural protection, seamlessly blending form and function [84]. To address feedback from P2, P5, P6, P9, P11, P12, P14, and P17 in User Study 2 on data readability and inspired by He et al. [43], we optimized the glowing effect to diffuse outward from the center. In the device layer of the Living Bento, white LEDs arranged in three concentric layers increase in number from the innermost to outermost layer. The LEDs light up sequentially based on detected resting heart rates (60–100 bpm) [17, 87, 88]: low rates illuminate the innermost layer, moderate rates the middle layer, and high rates the outermost layer (Figure 7). This dynamic lighting creates a layered visual experience reflecting users' heart rate changes. Lastly, the isolation layer adopts a double-layer design that ensures light transmission, blocks food liquids, secures food with cross-shaped perforations, and prevents light leakage to enhance the food's luminous effect.

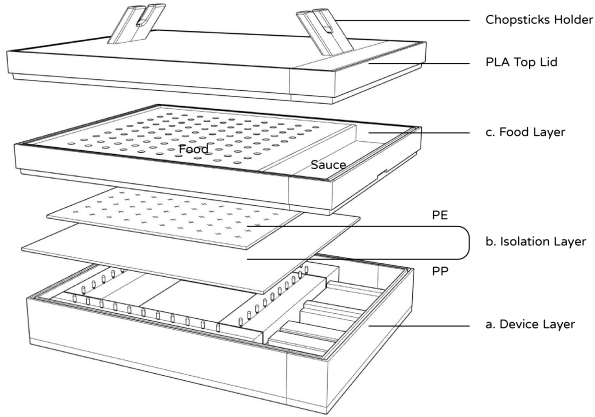

**Figure 6: Mechanical Design**

Chopsticks Holder
PLA Top Lid
c. Food Layer
PE
b. Isolation Layer
PP
a. Device Layer

---

[8]https://www.majimelife.com.au/collections/all-bento-boxes

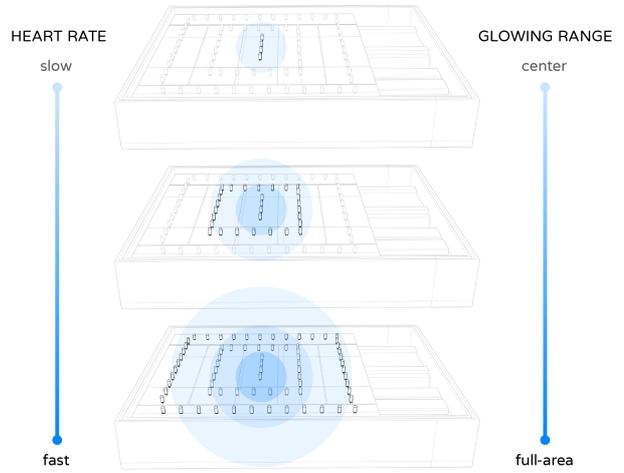

HEART RATE
slow

GLOWING RANGE
center

fast

full-area

**Figure 7: The Glowing Range according to the User's Heart Rate**

*5.3.2 User Experience.* In User Study 2, P5, P11, P15, P17, and P19 noted that the pulse sensor attached to their fingers caused inconvenience while eating. To address this, we integrated the "training chopsticks" concept with electronic components, embedding a pulse sensor into the chopsticks for real-time heart rate monitoring during dining (Figure 8), leveraging the familiar format and user-centered design principles [75, 83].

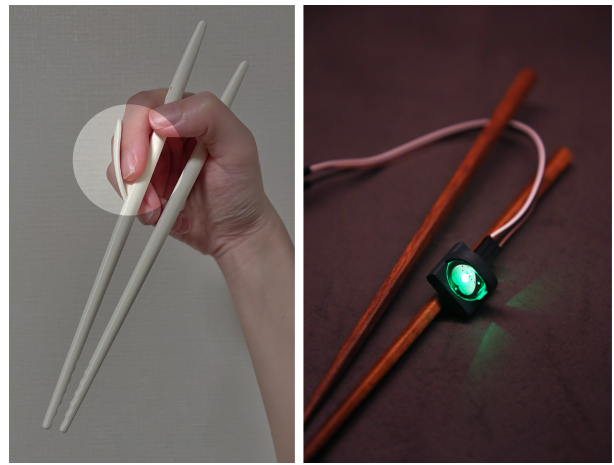

**Figure 8: The Comparison Between Conventional Training Chopsticks and Chopsticks with Pulse Sensor**

*5.3.3 Food Placement Design.* To strengthen the previously mentioned self-prioritization effect, we not only integrated diners' physiological data with the food but also focused on visual placement customization, as individuals prefer customized food [109]. Inspired by weaving techniques as an example of customization, we designed a 10 x 10 grid food layer, allowing users to weave various patterns based on their preferences using translucent ingredients. By utilizing a 2 x 2 weaving unit, users can create 2D patterns or adjust the height of the ingredients to form 3D structures (Figure 9). Based on feedback from P9 in User Study 2, who stated that any translucent ingredients could be used in this system, we demonstrated this design using translucent vermicelli noodles [81, 82]. The grid features evenly distributed holes sized to accommodate the vermicelli, producing a glowing effect resembling soft neon lights.

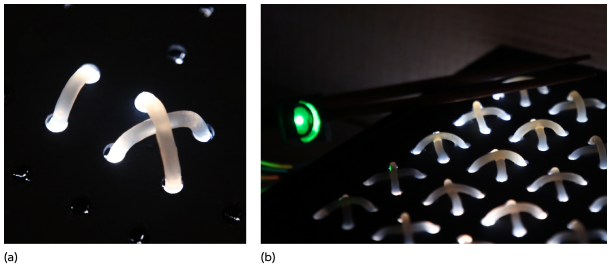

(a)                    (b)

**Figure 9: (a) 2 x 2 Weaving Unit (b) Different Weaving such as Crosshatch Pattern**

## 6 Limitation and Design Implications

### 6.1 Limitations

First, our experiments, workshops, and product development focused on specific translucent foods, such as Kudzu Jelly Noodles or vermicelli. Although other translucent foods [60, 142], like seaweed [32], agar agar [93], dumplings [36, 72], and rice paper [72], could potentially be used, our findings cannot be generalized to other food types without further research. Additionally, in User Study 1, we did not require fasting [61]. While it is not always necessary for dining experiences studies [13, 129], which could introduce variability due to hunger levels. Therefore, User Study 2 mandated fasting for one hour prior to participation. In addition, while our follow-up study compared synthesized and real-time heartbeat-driven glowing food to assess the effectiveness of the heartbeat-driven pattern, it was an initial exploration, and future research will require a larger sample size.

In User Study 2, we did not regulate intimacy levels between participant pairs, leading to variations in feedback. However, this finding suggests that familiarity influences the effectiveness of the heartbeat-sharing system in enhancing emotional connections. For friends with strong emotional bonds, sharing this data may not deepen their connection, while for less familiar individuals, it could foster stronger emotional interactions. Future research could explore the impact of relationship intimacy on the effectiveness of this system. Additionally, some participants expressed a desire to influence their partner's emotions through conversation, indicating potential risks of manipulation based on perceived emotional outcomes [26]. Future studies should address power dynamics among dining partners to mitigate these risks [26].

### 6.2 Design implications

*6.2.1 Bridging Cultural or Ethnic Differences.* The interviews from User Study 2 indicate that the "Sharing Bio-Sync Food" system has the potential to strengthen relationships that are initially less intimate. Therefore, inspired by Eat Love [117] and Sharing Dinner [139] by food designer Marije Vogelzang[9], we aim to use the Living Bento as a bridge between different cultures or ethnicities, facilitating connections between unfamiliar or even opposing cultures to alleviate barriers and foster positive beginnings. While we initially recruited participants who came with a companion, and though some pairs had low intimacy, they were at least acquainted with each other. Moving forward, we plan to recruit pairs from different cultures who do not know each other to assess the effectiveness of Living Bento in building friendship bridges across cultures.

*6.2.2 Development of Remote Dining Systems.* In User Study 2, some participants expressed that the heartbeat-sharing system provided a sense of companionship and reduced feelings of loneliness. Some even noted that, despite not having their dining partner physically present, the representation of their partner's heartbeat in their food gave them a feeling of their companion being nearby. Therefore, we foresee the potential of Living Bento to develop into a remote dining system. Although there is considerable research on remote dining systems in the HFI field [44, 134, 140], studies exploring remote dining through "Sharing Bio-Sync Food" remain limited. We will pursue further development in this area.

*6.2.3 Expanding Heartbeat-Driven Interaction Beyond Dining Experiences.* Building on the concept of resonance [14], biofeedback has been shown to influence physiological states and regulate emotions [18]. In User Study 2, P3, P15, and P20 reported feeling more relaxed when experiencing their partner's slower heart rate, suggesting that sharing heartbeat-driven feedback could potentially enhance emotional modulation and well-being in various contexts, from stress management to emotional support in collaborative settings.

*6.2.4 Impact of Awareness Duration.* Some participants mentioned that their emotional state stabilized over time during the eating process, shifting from initial novelty and excitement to a habituation of the blinking light. Future research should test the duration of awareness with Living Bento and clarify the relationship between the duration of awareness and the dining experience. We need to determine whether a gradual stabilization of emotions or sustained high levels of awareness provides a better dining experience.

*6.2.5 Optimization and Expansion of Living Bento.* Although a battery is integrated into the dining utensils, using wired pulse sensor chopsticks may be inconvenient for both diners and chefs. To improve user experience and flexibility, we aim to develop a wireless version of Living Bento with wireless communication modules (e.g., Bluetooth or Wi-Fi) in the next iteration. Furthermore, the structure of Living Bento will be reimagined into a more convenient utensil form, with efforts focused on developing alternative methods for broader generalizability.

[9]https://www.marijevogelzang.nl

## 7 Conclusion

In this study, we provided evidence that heartbeat-driven glowing food enhances various aspects of the dining experience. These findings address the negative effects of mindless eating and strengthen the relationship between individuals and food. Additionally, we developed a system of "Sharing Bio-Sync Food" to enhance emotional connections between individuals through food, and developed the Living Bento to implement the concept of "eating glowing heartbeats" in real-life settings. As Jean Anthelme Brillat-Savarin, often regarded as the first gastronome, famously stated: "Tell me what you eat and I shall tell you what you are" [11]. We have realized Brillat-Savarin's adage by implementing glowing food driven by one's own heartbeat, embodying the notion of "you are what you eat."

## Acknowledgments

The authors sincerely express their gratitude to Sohei Wakisaka for providing invaluable advice during this research and to Mingyang Xu for his active assistance with the follow-up study of User Study 1. Additionally, this work was supported by JST SPRING, Japan Grant Number JPMJSP2123. These contributions and support were instrumental in the successful completion of this research.